\documentclass[twocolumn,showpacs,preprintnumbers,amsmath,amssymb,epsfig]{revtex4}

\begin{document}

\title{Observational Evidence for Extra Dimensions from Dark Matter}

\author{Bo Qin$^{1}$ Ue-Li Pen$^{2}$ \& Joseph Silk$^{3}$}

\email{qinbo@bao.ac.cn, pen@cita.utoronto.ca, silk@astro.ox.ac.uk}

 \affiliation{
1. National Astronomical Observatories, Chinese Academy of Sciences, 
   Beijing, 100012, China\\
2. Canadian Institute for Theoretical Astrophysics, University of Toronto, 
   ON M5S 3H8, Canada\\
3. Astrophysics Department, University of Oxford, Oxford OX1 3RH, UK}

\begin{abstract}

Recent astronomical observations of systems of dark matter, which have been 
cited as providing possible support for self-interacting cold dark matter, 
may provide evidence for the extra dimensions predicted by superstring 
scenarios. We find that the properties of the required dark matter 
self-interaction are precisely the consequences of a world with 3 large 
extra dimensions of size $\sim$1nm, where gravity follows the $r^{-5}$ law 
at scales below $\sim$1nm. From the cross sections measured for various dark 
matter systems, we also constrain the mass of dark matter particles to be 
$m_x \sim 3\times 10^{-16}$ proton mass, consistent with the mass of 
axions.

\end{abstract}


\maketitle

String theory is a theory of all fundamental physical interactions. 
While in principle able to explain all phenomena, it has in practice been 
very challenging to find any observable predictions. One generic prediction 
is the existence of extra dimensions in addition to our familiar 
3-dimensional space. These extra dimensions had been thought to be extremely 
tiny (of order the Planck length $\sim 10^{-33}$cm), until in recent years 
the idea of large extra dimensions was proposed to address the 
``hierarchy problem''---a problem associated with the factor of the $10^{17}$  
huge difference between the Planck scale and the weak interaction scale. 
In the Arkani-Hamed, Dimopoulos and Dvali (ADD) scenario, the electro-weak 
and Planck energies are the same, and the large extra dimensions explain 
the apparent discrepancy in strength when measured on macroscopic 
scales \cite{ref1}. In the context of string cosmology, 
the fundamental scale would be $\sim 1$TeV, corresponding to $10^{-17}$cm. 
There are $3+1$ cosmological size dimensions, $n$ large extra dimensions, 
and $6-n$ fundamental scale dimensions.

The size $R$ of these large extra dimensions depends on the number of large 
extra dimensions $n$ . Gravity would start to deviate from Newton's 
inverse square law at small distance scales $r < R$. For $n=2$, $R\sim 1$mm. 
This opens a new window for experimental tests of string theory and searches 
for extra dimensions, by precise measurements of the gravitational force 
at sub-mm scales. Tremendous efforts have been made in the past few years 
in testing Newton's inverse square law at small scales and searching for 
evidence of the large extra dimensions. Currently, the measurements are 
reaching micron scales and no deviation from Newton's law 
has been found from $\sim$1cm down to $\sim 10^{-3}$cm 
\cite{ref2,ref3,ref4,ref5}.

While string scenarios and extra dimensions have not yet been tested 
experimentally so far in the laboratory, recent astronomical observations 
of dark matter, on the other hand, may shed light on the issue. It should 
be noted that any connection of string scenarios to observable phenomena 
would be an exciting possibility deserving further investigation.

The existence of dark matter was first realized by Zwicky nearly 70 years 
ago \cite{ref6}, and it is now well established that dark matter constitutes the 
major matter component of the universe \cite{ref7}. Although its nature 
remains a mystery, most cosmologists and particle physicists believe that 
dark matter is likely to be a new species of elementary particle that 
is neutral, long-lived, cold (or nonrelativistic), and collisionless 
(i.e. dark matter particles have very little interactions with themselves 
as well as with ordinary matter). This ``standard'' picture of collisionless 
cold dark matter (CCDM) has gained great success in explaining the origin 
and evolution of cosmic structures on large scales \cite{ref7,ref8,ref9,ref10}.

However, the CCDM model is facing a potential challenge in recent years 
from observations on galactic and sub-galactic scales \cite{ref8,ref11}. 
Numerical simulations of the CCDM model predict that the density profiles 
of dark matter halos should exhibit a cuspy core in which the density 
rises sharply as the distance from the center decreases \cite{ref12}. 
In contrast, observations of systems of dark matter, ranging from 
dwarf galaxies \cite{ref13,ref14,ref15,ref16}, 
low surface brightness galaxies \cite{ref17,ref18} to galaxies 
comparable in mass with the Milky Way \cite{ref19}, indicate that 
the central density profiles are probably much less cuspy than predicted. 
Clusters of galaxies also seem to reveal a near isothermal core 
\cite{ref20,ref21}, although there is considerable scatter 
\cite{ref22}.

The plausible discrepancies between theory and observations, although 
still vigorously debated, have stimulated many attempts to understand 
the nature of dark matter and to modify the CCDM model, among which one of 
the more popular schemes is the self-interacting cold dark matter model. 
As proposed by Spergel and Steinhardt, the above conflicts can be 
readily resolved if the cold dark matter particles are self-interacting 
with a large scattering cross section 
$\sigma_{xx}/m_x \approx 8 \times 10^{-(25-22)}$cm$^2$/GeV, 
where $m_x$ is the mass of dark matter particles \cite{ref11}.

Follow-up work by many authors, however, suggests that the issue 
may be more complicated. In galaxy clusters, dark matter was found to 
probably have a much weaker self-interaction. Gravitational lensing 
studies of clusters have placed an upper limit
$\sigma_{xx}/m_x < 10^{-25.5}$cm$^2$/GeV \cite{ref23}. 
Detailed X-ray observations of clusters revealed 
that 
$\sigma_{xx}/m_x < 2 \times 10^{-25}$cm$^2$/GeV \cite{ref24,ref25}, 
and a joint X-ray/weak-lensing study showed that 
$\sigma_{xx}/m_x < 2 \times 10^{-24}$cm$^2$/GeV \cite{ref26}. 
From numerical simulations, Yoshida {\it et al.} found that the cross 
sections needed to produce good agreement with galaxies turned out to 
produce galaxy cluster cores that were too large and too round to be 
consistent with observation \cite{ref27}.

These lines of evidence might suggest that, instead of been a fixed 
value, the dark matter self-interaction cross section as proposed by 
Spergel and Steinhardt, is likely to vary in different dark matter 
systems---smaller systems (like dwarf galaxies) tend to have larger 
cross sections whereas more massive systems (like galaxy clusters) 
tend to have smaller cross sections. 
Indeed, Firmani {\it et al.} have constrained 
the cross section from various dark matter systems ranging from dwarf 
galaxies to galaxy clusters, and proposed a relation \cite{ref28}
\begin{equation}
\frac{\sigma_{xx}}{m_x} \approx 4 \times 10^{-25}
           \left( \frac{100 \; {\rm km} \, {\rm s}^{-1} } {v}
           \right) \ \ \rm{cm}^2/\rm{GeV}
\end{equation}
where $v$ is the velocity dispersion of the dark matter system. 
Note that more massive systems (like galaxy clusters) have higher 
velocity dispersions while less massive systems (like dwarf galaxies)
have lower velocity dispersions.

The nature of this self-interaction between dark matter particles is unknown. 
Its strength generally must be put in by hand. We note that the scattering 
cross section in Eq.~(1) decreases with increasing velocity, which is 
characteristic of long-range forces (like gravity or Coulomb forces). 
On the other hand, as addressed by ADD, gravity has only been accurately 
measured in the $\sim$1cm range but has been extrapolated to small distance 
scales on the assumption that gravity is unmodified over the 33 orders of 
magnitude from $\sim$1cm down to the Planck length $\sim 10^{-33}$cm. 
Will gravity still follow the conventional Newtonian $r^{-2}$ law at 
extremely small distance scales? Does the dark matter self-interaction 
have anything to do with the (microscopic) asymptotic behavior of gravity?

Here we propose that if gravity deviates from Newton's inverse square law 
at sub-mm scales and varies as $r^{-(2+n)}$, as suggested by ADD, 
then the strength of gravity would be greatly enhanced at small distance 
scales and hence could {\it naturally} provide the self-interaction for 
dark matter particles, without introducing a new interaction which would 
otherwise seriously complicate the Standard Model. The self-interaction 
between dark matter particles, if true, may have strong implications 
for the modification of gravity at small distance scales that 
experimental physicists have been searching for during the past few years. 
Also, from the cross sections determined for various dark matter systems, 
we can constrain how gravity varies with distance.

We assume that at small distance scales below the ``critical radius'' $R$, 
gravity starts to deviate from the Newtonian $r^{-2}$ law and takes 
the general form:
\begin{equation}
F = \alpha \frac{G M m}{r^{2+n}},
\end{equation}
where $\alpha$ is a constant with dimension [length]$^n$, 
$G$ is the gravitational constant, 
$M$ and $m$ are the masses of the two particles. 
The value of $\alpha$ is easily determined to be $\alpha=R^n$,
from the boundary condition that at $r=R$, 
$\alpha \frac{G M m}{r^{2+n}} = \frac{G M m}{r^2} $,

When dark matter particles with a mean relative velocity $u$ approach 
close enough to each other, gravitational scattering may cause large 
deflection angles. 
For particles moving in the central force field described by Eq.~(2),
the elastic scattering cross section is given by \cite{ref29}
\begin{equation}
\sigma = \pi A \left( \frac{\alpha G m_x}{u^2}  \right)^{\frac{2}{n+1}}, 
\end{equation}
where
$A= [(n+1)/(n-1)]^{(n-1)/(n+1)}$ for $n>1$,
and 
$A=1$ for $n=1$.
The value of $A$ is close to $1.4$ for $2\leq n \leq 5$. 
Taking into account the relation $u^2 = 2 v^2$ between the mean 
relative velocity and the velocity dispersion, we can then estimate 
the elastic scattering cross section between dark matter particles as:
\begin{equation}
\sigma_{xx} = \pi A \left( \frac{\alpha G m_x}{2 v^2} 
                      \right)^{\frac{2}{n+1}}, 
\end{equation}
where $A=1$ for $n=1$, and $A \approx 1.4$ for $2\leq n \leq 5$. 
Combining Eqs~(1) and (4) we find that there is only one solution 
which is the $n=3$ case, corresponding to 3 large extra dimensions.
The case of $n\neq 3$ was excluded, because Eqs~(1) and (4) 
result in an unrealistic solution where the value of $m_x$ 
varies with $v$.

The above calculation of the elastic scattering cross section was 
based on classical mechanics. However, for the problem discussed
in this letter, the de Broglie wavelength of the particles is 
much greater than the length scale at which the particles 
interact with each other. This can be seen from the smallness 
of the dark matter particle mass (shown in following paragraphs). 
Hence, a quantum mechanical treatment is required. 
Fortunately, for the elastic scattering cross section in the central 
force field as in our case, quantum mechanics gives similar results 
as classical mechanics.

From a detailed calculation, Vogt and Wannier
have shown that the quantum mechanical scattering cross section 
is exactly twice the classical value, for a central force field
of the $r^{-5}$ form \cite{ref30}, corresponding to our $n=3$ case. 
For the general case of $n$, Joachain has shown that 
the boson-boson identical particle scattering cross section 
in s-wave is twice the classical cross section, while the s-wave 
scattering cross section of identical fermions is four times
smaller than the boson-boson scattering \cite{ref31}. Note that 
our case is in the low energy regime where the s-wave scattering 
dominates. Therefore, our classical constraints on $n$ from 
Eqs~(1) and (4) are still valid in the quantum mechanical regime.
A factor of a few difference in $\sigma_{xx}$ does not change 
the value of $n$ we have derived above.

Correcting Eq.~(4)
by a factor of two, we have the quantum mechanical
cross section in the $n=3$ case:
\begin{equation}
\sigma_{xx} = 2 \pi \frac{ (\alpha G m_x)^{\frac{1}{2}} }{v} 
                    , \ \ \ \  {\rm for} \  n=3.
\end{equation}
Combining Eqs~(1) and (5) we obtain  
$m_x= 3\times 10^5 \alpha$~GeV, with $\alpha = R^n$. 
According to ADD, the size $R$ of the extra dimensions can be
expressed as $R \sim 10^{\frac{30}{n}-17}$cm. 
We therefore find that $R\sim 10^{-7}$cm and the dark matter 
particle mass $m_x \sim 3\times 10^{-16}$GeV.

By attributing the dark matter self-interaction to modified gravity at 
small distance scales as suggested in the ADD scenario, we have obtained 
the following results:

(1) We have avoided the introduction of a new fine-tuned interaction in 
the Standard Model by using an existing physical scenario.

(2) The number of large extra dimensions has been constrained and found 
to be $n=3$. 

(3) The size of the large extra dimensions was found to be of order 
$R \sim 10^{-7}$cm, below which gravity deviates from Newton's inverse 
square law and varies as $r^{-5}$.

(4) The mass of dark matter particles was constrained to be 
$m_x \sim 3\times 10^{-16}$ proton mass---falling into roughly
the mass range of the axion which has been proposed as a dark matter 
candidate (see e.g. \cite{ref32}).

This ties together the large extra dimension string scenario proposed 
solely for particle physics and the dark matter halo structure puzzle in 
astrophysics, without introducing any new or fine-tuned parameters.  
The standard self-interacting dark matter solution to the problem 
requires fine-tuned cross sections, and does not explain the scale-dependent 
behavior. Interaction cross sections are enhanced over the characteristic 
electro-weak values because of the relatively slow motions of particles 
in dark matter halos.

It should be noted that the issue of self-interacting dark matter 
is still under debate. In more massive systems like galaxy clusters, 
the self-interaction cross section appears to be small, while in 
less massive systems like low surface brightness galaxies and 
dwarf galaxies, the larger cross section could be due to 
mis-interpretation of the observed data. Further studies
are needed to confirm the self-interacting dark matter model.
The $n$ and $m_x$ values we obtained from the empirical formula
Eq.~(1) is sensitive to the form of cross section determined 
from observations. A cross section different 
from Eq.~(1) will result in different $n$ and $m_x$: 
the dependance of $\sigma_{xx}$ on the velocity dispersion $v$ 
determines the value of $n$, and $m_x$ is also affected by
the coefficients in Eq.~(1). Furthermore, $m_x$ is sensitive 
to the exact form of $R$ as a function of $n$ given by ADD. 
Overall, our constraints on the extra dimensions relies on an 
accurate determination of the self-interaction cross section.

Nevertheless, we are suggesting a potentially new connection between 
string scenarios and astrophysical phenomena. The ADD scenario with 
3 large extra dimensions naturally explains the velocity-dependent 
cross section (as in Eq.~(1)) for self-interacting dark matter, 
and predicts the number 
and size of extra dimensions, and also the dark matter particle mass.  
The current limits on gravity at small scales are problematic for the 
original $n=2$ proposal in the ADD scenario, but our model predicts 
deviations from Newtonian gravity at the nanometer scale. The generic 
prediction of the ADD scenario is a wealth of quantum gravity phenomena 
at the LHC.

Ordinarily, string scenarios are evaluated on the basis of aesthetics and 
heuristic mathematical arguments.  For example, it has been speculated that 
the large extra dimensions might have instabilities. The empirically 
appealing aspect of the ADD scenario is its connection to observable 
phenomena, which allows a scientific evaluation, via test or falsification.

We thank Xuelei Chen, Carlos S. Frenk, Yannick Mellier, Xiang-Ping Wu and 
Hongsheng Zhao for comments and discussions. B.Q. is grateful to Ya-Ping Zhang 
for help and the National Science Foundation of China for support under 
grant 10233040.

\end{document}